\documentclass[twocolumn, preprintnumbers,amsmath,amssymb]{revtex4}

\usepackage{graphicx}
\usepackage{dcolumn}
\usepackage{bm}
\usepackage{color}

\begin{document}

\preprint{}
                           
\title{Site-Selective Field Emission Source by Femtosecond Laser Pulses \\and Its Emission Mechanism}

\author{Hirofumi Yanagisawa}
\email{hirofumi@phys.ethz.ch}
\affiliation{Department of Physics, ETH Z\"{u}rich, Wolfgang-Pauli-Strasse 16, CH-8093 Z\"{u}rich, Switzerland}

\begin{abstract}
Recent experimental and theoretical investigations on asymmetric field emission induced by weak femtosecond laser pulses and also its emission mechanisms are briefly reviewed. The emission mechanisms are discussed further for a wider range of DC fields and laser power. It appears that firstly photo-assisted field emission from lower-excitation order grows in the higher DC fields and secondly our simulations can be applied only for lower laser power.
\end{abstract}

\date{\today}
\maketitle

\section{Introduction}
Field emission is the electron emission from a solid to the vacuum via electron tunneling through the potential barrier at the surface. It is driven by applying a strong electric field (2-5 V/nm) to the solid \cite{gomer93, fursey03}. Field emission from metallic tips with nanometer sharpness has been particularly interesting because its electron beam is highly coherent and bright due to a nanometer- or atomically- sized emission area, and it has been widely used in both basic research and applications \cite{nagaoka98, oshima02,cho04,fink86,fink90,fu01}. 

Field emission from such a tip can be induced with a moderate DC field and laser irradiation onto the tip apex. So called laser-induced field emission was introduced some time ago, first by using a continuous laser beam \cite{lee73} and later by pulsed lasers \cite{boussoukaya87, lee89}. More recently, ultrashort pulsed field emission by using femtosecond laser pulses was realized \cite{hommelhoff06a, hommelhoff06b}. The pulsed field emission can be emitted from almost the same emission area as that of field emission \cite{hommelhoff06a}.  Hence potentially a spatio-temporal resolution down to the single atom and attosecond range appears to be possible \cite{fink86,fu01,hommelhoff06a, hommelhoff06b}. Moreover, due to plasmonic effects, emission patterns were found to be asymmetric and can be controlled by laser polarization. In effect, an ultrafast pulsed field-emission source with site selectivity on the scale of a few tens of nanometers was realized \cite{yanagisawa09,yanagisawa10}. So far, the emission mechanisms have been intensively investigated for weak and strong field regimes \cite{hommelhoff06a,hommelhoff06b,yanagisawa09,yanagisawa10,ropers07,barwick07,wu08,tsujino09,yanagisawa11,hommelhoff10,ropers11,hommelhoff11,hommelhoff12a,hommelhoff12b,hommelhoff12c, ropers12}. Such electron sources would be very attractive for both basic research and new applications like time-resolved electron microscopy, spectroscopy, holography, and also free electron lasers \cite{hommelhoff06a,yanagisawa09, aeschlimann07, tonomura87, ropers07, ganter08, tsujino09}.

The laser-induced field emission is considered to be driven through the following two steps. 1. Illuminating the metallic tip with the laser pulses creates enhanced optical electric fields at the tip apex due to plasmonic effects (\emph{local field enhancement}) \cite{novotny97, novotny02, hecht05}. 2. The enhanced fields induce pulsed field emission in combination with a moderate DC voltage applied to the tip. Depending on the strength of the enhanced fields, different field emission mechanisms become dominant \cite{hommelhoff06a,hommelhoff06b,yanagisawa09,yanagisawa10,ropers07,barwick07,wu08,tsujino09,yanagisawa11,hommelhoff10,ropers11,hommelhoff11,hommelhoff12a,hommelhoff12b,hommelhoff12c,ropers12}.  For relatively weak fields, single-electron excitations by single- and multi-photon absorption are prevalent, and photo-excited electrons are tunneling through the surface potential barrier (\emph{photo-assisted field emission}) or emitted over the barrier (\emph{photoemission}). On the other hand, very strong fields largely modify the tunneling barrier and prompt field emission from the Fermi level (\emph{optical field emission}). 

In this manuscript, our recent investigation on the optical control of field emission sites and its emission mechanisms are briefly summarized in sections 2 and 3, respectively. Optical control of field emission sites was realized because the above mentioned local field distribution on the tip apex becomes asymmetric, thus driving the asymmetric field emission  \cite{yanagisawa09,yanagisawa10}. Energy distribution curves (EDCs) of the emitted electrons revealed that their emission mechanism is mainly dominated by photo-assisted field emission  \cite{yanagisawa11}. In section 3, we extend our discussion on EDCs for a wider range of DC fields and laser power. The results show a change of the emission processes and the spectral shapes with varying both parameters, and the validity of our simulations for EDCs appears to be limited to lower laser power.


\section{Optical control of field emission sites}

\begin{figure}[b!]
  \includegraphics[scale=0.25]{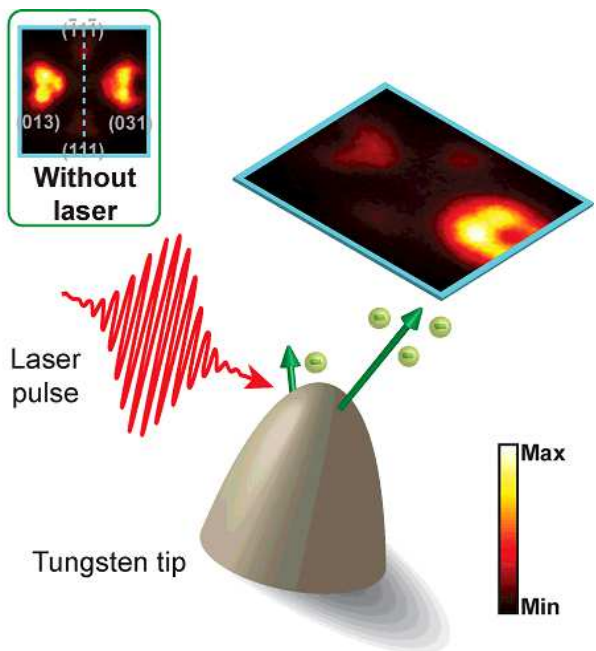}
  \caption{\label{fig:epsart}
   A schematic diagram of the laser-induced field emission. The emission pattern is observed by a two-dimensional detector. The inset shows the field emission pattern without laser excitation.}
\end{figure}

In our previous work, we have investigated electron emission patterns induced by femtosecond laser pulses from a clean tungsten tip apex, and compared them with those of field-emitted electrons without laser excitation as schematically depicted in Fig. 1. A tungsten tip with its axis along the (011) crystal direction is mounted inside a vacuum chamber. Laser pulses are generated in a Ti:sapphire oscillator (wave length: 800 nm; repetition rate: 76 MHz; pulse width: 55 fs; average laser power $P_{L}$: 20 mW) and introduced into the vacuum chamber. The laser light was focused to 4 $\mu m$ in diameter onto the tip apex, and emitted electrons were detected by a two-dimensional detector. The polarization vector can be changed within the transversal (x, z) plane by using a $\lambda/2$ plate. The polarization angle $\theta_{P}$ is defined by the angle between the tip axis and the polarization vector; details are depicted elsewhere \cite{yanagisawa09, yanagisawa10}.

\begin{figure}
  \includegraphics[scale=0.22]{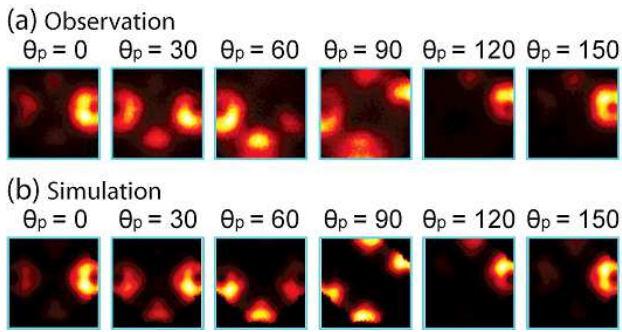}
  \caption{\label{fig:epsart}
    (a) Observed and (b) simulated laser-induced field emission images for different light polarization angles $\theta_{P}$. $\theta_{P}$ is defined by the angle between the tip axis and the polarization vector: $\theta_{P}=0$ is when the tip axis is parallel to the polarization vector.}
\end{figure}

We observed a striking difference in symmetry of the two patterns. Without laser, a typical field emission pattern of a clean W tip was observed. The most intense electron emission was observed around the (310)-type facets and relatively week emission from (111)-type facets as indicated in the picture. The intensity map roughly represents a work function map of the tip apex: the lower the work function is, the more electrons are emitted \cite{sato80,kittel,yanagisawa10}. On the other hand, with laser irradiation, emission sites were the same as those without laser, but the emission pattern becomes strongly asymmetric with respect to the shadow and exposed sides of the tip to the laser pulse. Furthermore, varying the laser polarization angle changes these distributions substantially as shown in Fig. 2(a). In effect, we have realized an ultrafast pulsed electron source with emission site selectivity of a few ten nanometers. Note that the emission site can be also selected by changing the azimuthal and the polar orientation of the tip apex relative to the laser incidence direction \cite{yanagisawa09, yanagisawa10}.

\begin{figure}[b!]
  \includegraphics[scale=0.22]{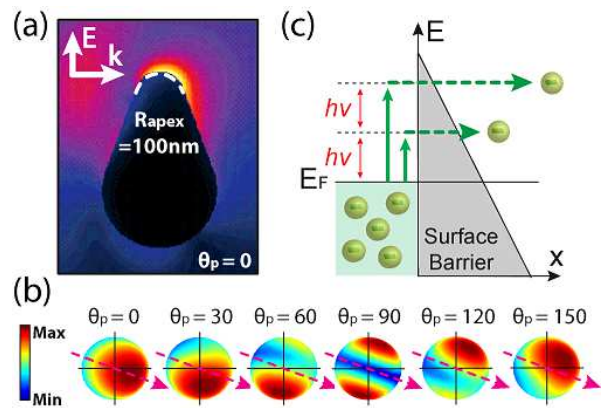}
  \caption{\label{fig:epsart}
The calculated time-averaged field distribution around the model tip is shown in a linear color scale in (a) for $\theta_{P}=0^\circ$. Highest field values are represented in yellow. In (b) the time-averaged field distributions are given in a front view of the model tip for different polarization directions $\theta_{P}$. The laser propagation direction is indicated by red arrows. (c) shows a schematic diagram of photo-assisted field emission from a nonequilibrium electron distribution..}
\end{figure}

The physics behind driving asymmetric field emission is the asymmetric local field distribution on the tip apex. When a laser pulse illuminates the metallic tip, surface electro-magnetic (EM) waves like surface plasmon polaritons are excited~\cite{zenneck}, which propagate around the tip apex. Due to the resulting interference pattern, the electric fields show an asymmetric distribution over the tip apex. To simulate the propagation of these surface EM waves and the resulting field distributions, we used the software package MaX-1 for solving Maxwell equations based on the Multiple Multipole Program~\cite{christian90, christian98, christian99}. A droplet-like shape was employed as a model tip as shown in Fig. 3(a), with a radius of curvature of the tip apex of 100 nm, which is a typical value for a clean tungsten tip. The dielectric function of tungsten at 800 nm was used \cite{tungstenepsilon}. 

Fig. 3(a) shows the calculated time-averaged field distribution over a cross section of the model tip, where the polarization vector has been chosen parallel to the tip axis ($\theta_{P}=0^\circ$). The laser propagates from left to right. The field distribution is clearly asymmetric with respect to the tip axis, with a maximum on the shadow side of the tip. This is consistent with our observations in Fig. 1 where the field emission is enhanced on the shadow side. Fig. 3(b) shows front views of time-averaged field distribution maps for different polarization angles. The field distribution changes strongly depending on the polarization angle. 

From the calculated local fields, we further simulated the laser-induced field emission images by supposing the photo-assisted field emission mechanism as schematically depicted in Fig. 3(c). The current density of field emission can be described in the Fowler-Nordheim theory based on the free-electron model~\cite{gomer93,fursey03, murphy56,young59}. The simulated emission patterns in Fig. 2(b) are in excellent agreement with the observations shown in Fig. 2(a). This comparison clearly demonstrates that the observed strongly asymmetric features originate from the modulation of the local photo-fields at the tip apex \cite{yanagisawa09, yanagisawa10}.

\section{Mechanism of laser-induced field emission}

\subsection{EDCs of field emission and photo-assisted field emission}
\begin{figure}[b!]
  \includegraphics[scale=0.2]{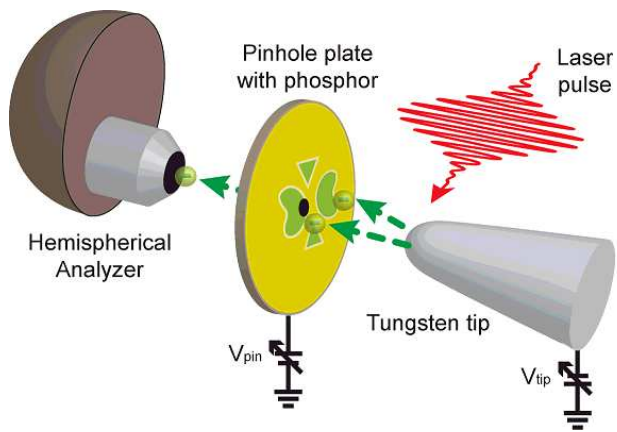}
  \caption{\label{fig:epsart}
Schematic diagrams of the experimental setup. The emission pattern can be observed at the phosphor plate as indicated by green areas. An emission site is selected by the pinhole (black dot) and EDCs are measured by a hemispherical analyzer. $V_{pin}$ and $V_{tip}$ represents the pinhole-plate voltage and tip voltage, respectively.}
\end{figure}

In the previous section, the photo-assisted field emission is shown to simulate well laser-induced field emission patterns. In order to confirm experimentally the emission mechanism, we upgraded our previous field emission setup by implementing an electron energy analyzer (see Fig. 4) and measured EDCs of the emitted electrons. In order to distinguish electrons emitted from different facets of the tungsten tip, a plate with phosphor coating and pinhole in the center was installed as an aperture and counter electrode between the tip and the analyzer. The tip and the pinhole plate can be negatively and positively biased for field emission, respectively. The field emission pattern of the clean tungsten tip can be observed on the phosphor plate where the most intense electron emission come from (310)-type facets. The pinhole is positioned at the edge of a (310) type facet as in Fig. 4. Note that the selected site is also the most intense emission site in the laser-induced field emission because linearly polarized laser light with the polarization vector parallel to the tip axis was used \cite{yanagisawa09, yanagisawa10}.


\begin{figure}[t!]
  \includegraphics[scale=0.22]{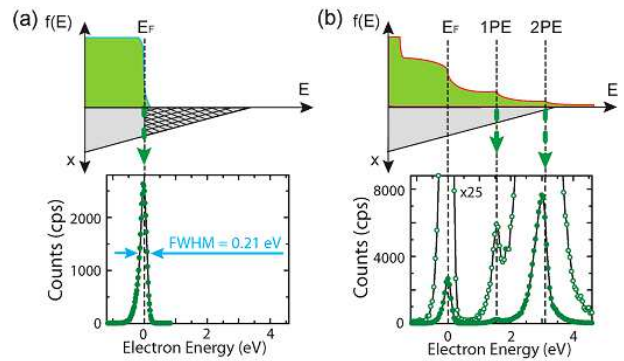}
  \caption{\label{fig:epsart}
(a) EDC of field emission from the tungsten tip together with a schematic diagram of field emission from a Fermi-Dirac distribution tunneling through a surface barrier. The transmission probability depends on the hatched area. The tip voltage $V_{tip}$ was -2300 V, and the pinhole-plate voltage $V_{pin}$ was 0 V. (b) shows experimental EDCs and schematic diagrams of photo-assisted field emission together with the magnified spectrum. The spectrum is taken at $V_{tip}$ = -2300 V, $V_{pin}$ = 0 V and the laser power $P_{L}$ = 50 mW. }
\end{figure}

Fig. 5(a) shows an EDC of field emission at a tip voltage $V_{tip}$ of -2300 V and a pinhole-plate voltage $V_{pin}$ of 0 V. The peak of the spectrum is defined as the Fermi energy $E_{F}$ at 0 eV \cite{comment1}. The spectrum shows a typical asymmetric peak, which can be understood by the diagram in Fig. 5(a). The field emission current is influenced by two factors: 1) the electron occupation number and 2) the transmission probability through the surface barrier \cite{gomer93,fursey03,murphy56,young59}. The occupation number is given by an electron distribution function $f(E)$, which is the Fermi-Dirac distribution function in the case of field emission, and the transmission probability depends exponentially on an area of the surface barrier above an energy of the emitted electron indicated by the hatched area. Therefore, the positive energy side of the spectrum falls off due to a rapid decrease of the occupation number, while the negative energy side falls off because of the exponential decay of the transmission probability due to the increase of the surface barrier area. Thus a typical field emission spectrum shows such an asymmetric peak. 

Upon low-intensity laser irradiation, the electron distribution is modified by single-electron excitations due to multi-photon absorption, resulting in a nonequilibrium distribution characterized by a step-like profile as illustrated in the inset of Fig. 5(b) \cite{wu08,lisowski04,rethfeld02}; the width of each step corresponds to the photon energy $h\nu$ (= 1.55 eV). Here we identify the step edges of one-photon excitation (1PE) and 2PE as shown in the inset. In a real situation, the excited electrons relax mainly by electron-electron ({\it e-e}) scattering on a time scale of a few femtoseconds, and such an electron dynamics results in a smeared electron distribution as in the diagram. These features should be reflected in the EDCs.

Fig. 5(b) shows an EDC of laser-induced field emission at $V_{tip}$ = -2300 V and a laser power $P_{L}$ of 50 mW. The spectrum shows the field emission peak undisturbed with identical shape and intensity as in Fig. 5(a), and additional peaks at the 1PE and 2PE edges are clearly observed. The latter show the same asymmetric shape as the field emission peak. Thus, photo-assisted field emission is confirmed experimentally. Regarding the relative intensities, photo-assisted field emission from 2PE is much stronger than that from 1PE even though the occupation number at 1PE is higher; this is because the transmission probability at 2PE is quite high. Note that photo-assisted field emission from 2PE has not been observed for excitation with a continuous-wave laser \cite{lee73}: those electron distributions are supposed to be largely different from our case.


\subsection{DC field and laser power dependences of EDCs}

\begin{figure}[b!]
  \includegraphics[scale=0.32]{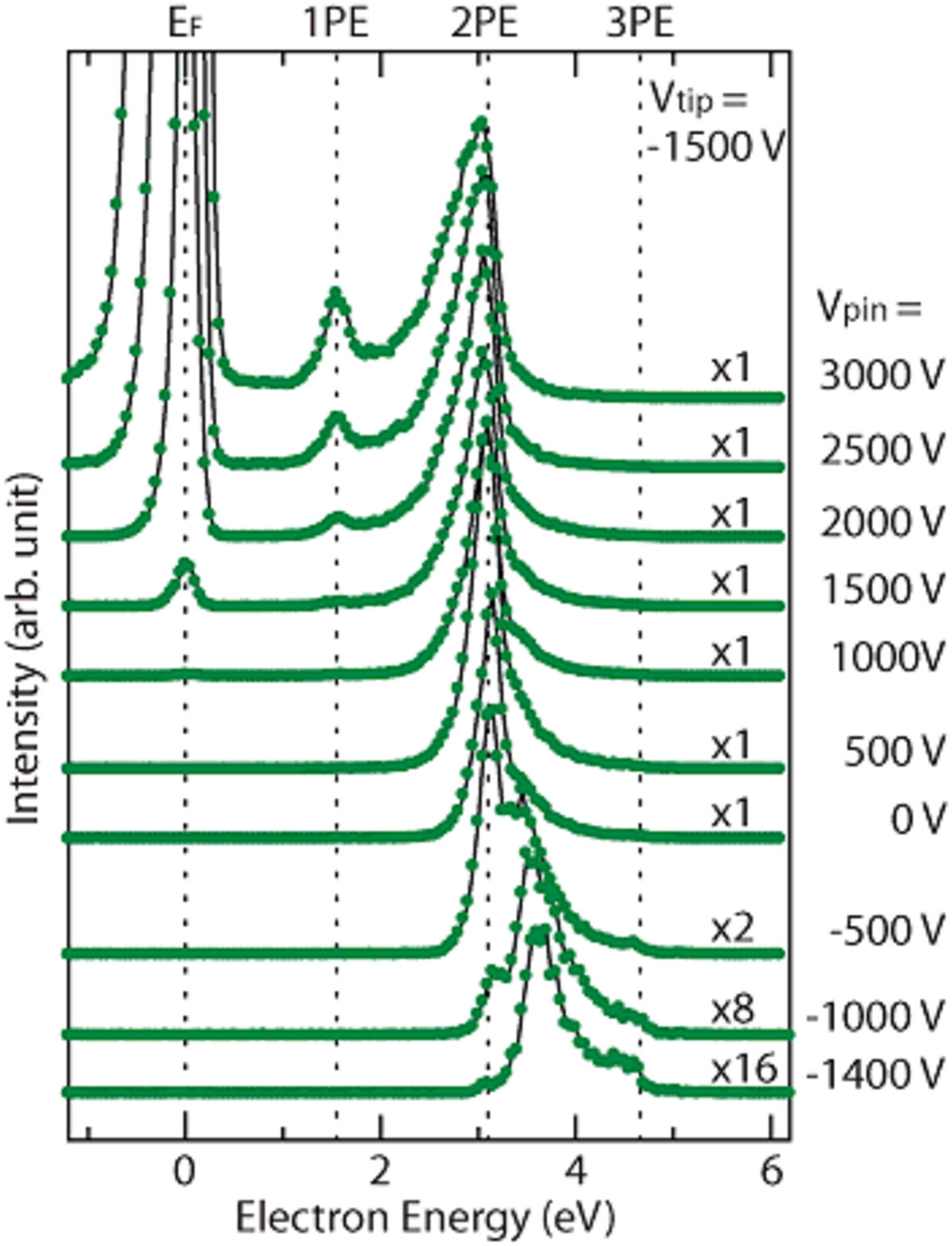}
  \caption{\label{fig:epsart}
Various EDCs for different $V_{pin}$ from -1400 V to 3000 V. $V_{tip}$ is set at -1500 V. Magnification is also shown with each spectrum. }
\end{figure}

\begin{figure*}[t!]
  \includegraphics[scale=0.32]{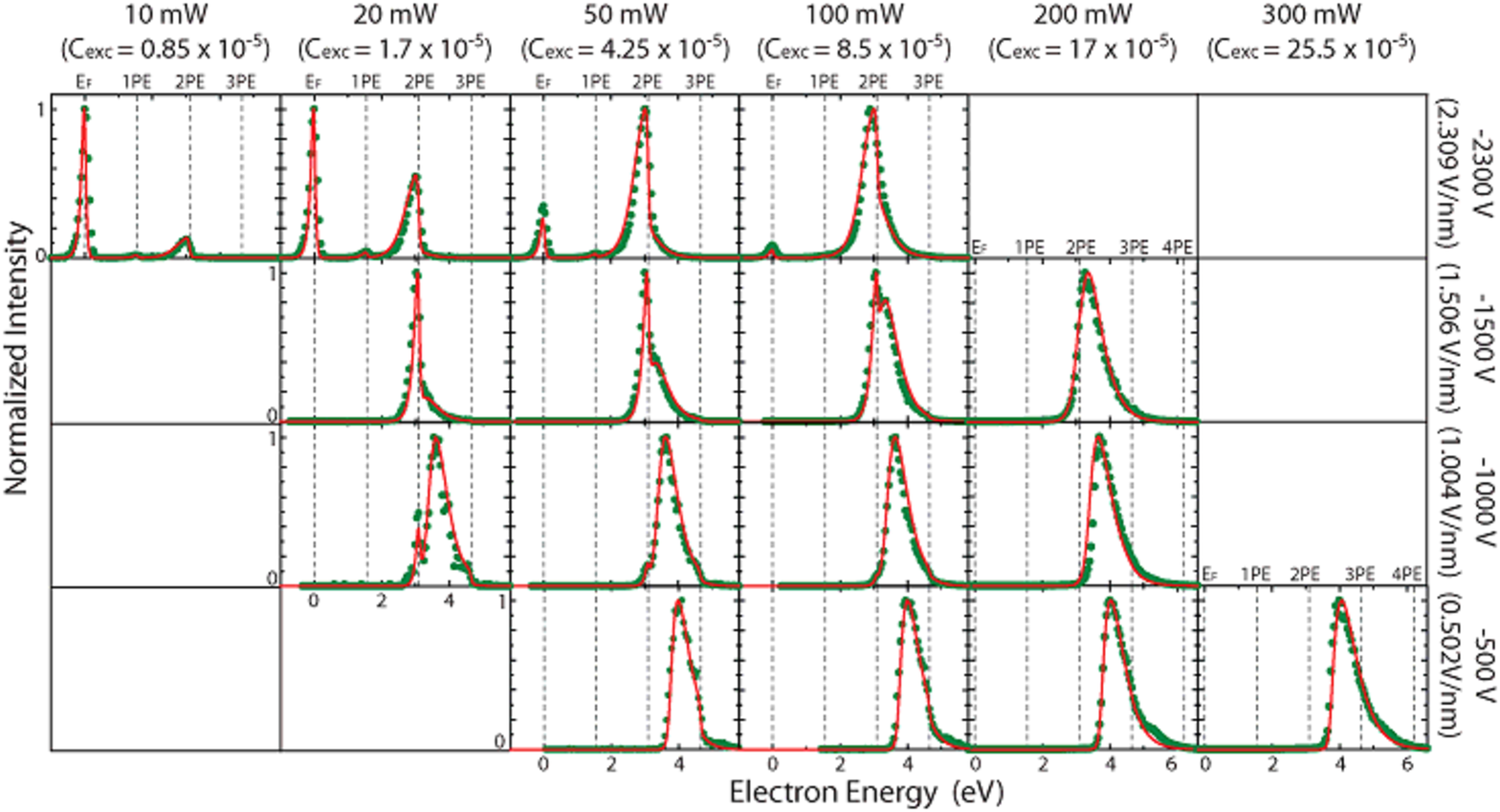}
  \caption{\label{fig:epsart}
Measured and simulated EDCs for different tip voltages (DC fields) and laser powers (excitation constant $C_{exc}$; appears in eq. (1) of Ref. \cite{yanagisawa11}).  The work function $\Phi$ of 4.592 eV was used for the simulation. All the spectra are normalized at their maximum values. The measured EDCs are shown as green dots, and the simulated EDCs as red lines. }
\end{figure*}

The emission processes and the spectral shapes can be tuned by changing the DC fields. Fig. 6 shows various EDCs for different DC fields, where the EDCs are taken with setting $V_{tip}$ at -1500 V and varying $V_{pin}$ from -1400 V to 3000 V while $P_{L}$ was fixed at 50 mW. These spectra show a smooth parametric transition between field emission, photo-assisted field emission and photoemission. In the lower voltages (high negative), the tunneling probability becomes suppressed and the emission spectrum is dominated by photoelectrons excited in a three-photon process as shown in the bottom spectrum of Fig. 6. The peak shows a spectral shape completely different from that of field emission. The peak maximum is located at approximately 0.65 eV below the 3PE edge. Since the transmission probability is unity throughout the photoemission regime, the peak shape reflects more closely the electron distribution function. Therefore the spectral shape indicates a strong modulation of the electron distribution due to {\it e-e} scattering processes. For the higher voltages, the tunneling probability becomes larger and  the photo-assisted field emission process becomes dominant. At around $V_{pin}$ = 1000 V, photo-assisted field emission from 2PE is dominant, and at higher $V_{pin}$, photo-assisted field emission from 1PE starts to grow \cite{comment2}.

The emission processes and the spectral shapes can also be tuned by changing the laser power. The laser power dependence of the EDCs was investigated by systematically measuring EDCs for various tip voltages and laser powers as shown in Fig. 7. These spectra are normalized at their maximum intensities to focus on their spectral shapes. In this case,  $V_{pin}$ is grounded and only $V_{tip}$ is varied. As a rule of thumb, with increasing laser power, the electron emission from higher order excitation becomes dominant, and the spectra become broader and smeared out, which holds for each voltage setting. The reason why the spectra become smeared out with increasing laser power is that more excited electrons can scatter into lower energy states.


\begin{figure*}[t!]
  \includegraphics[scale=0.30]{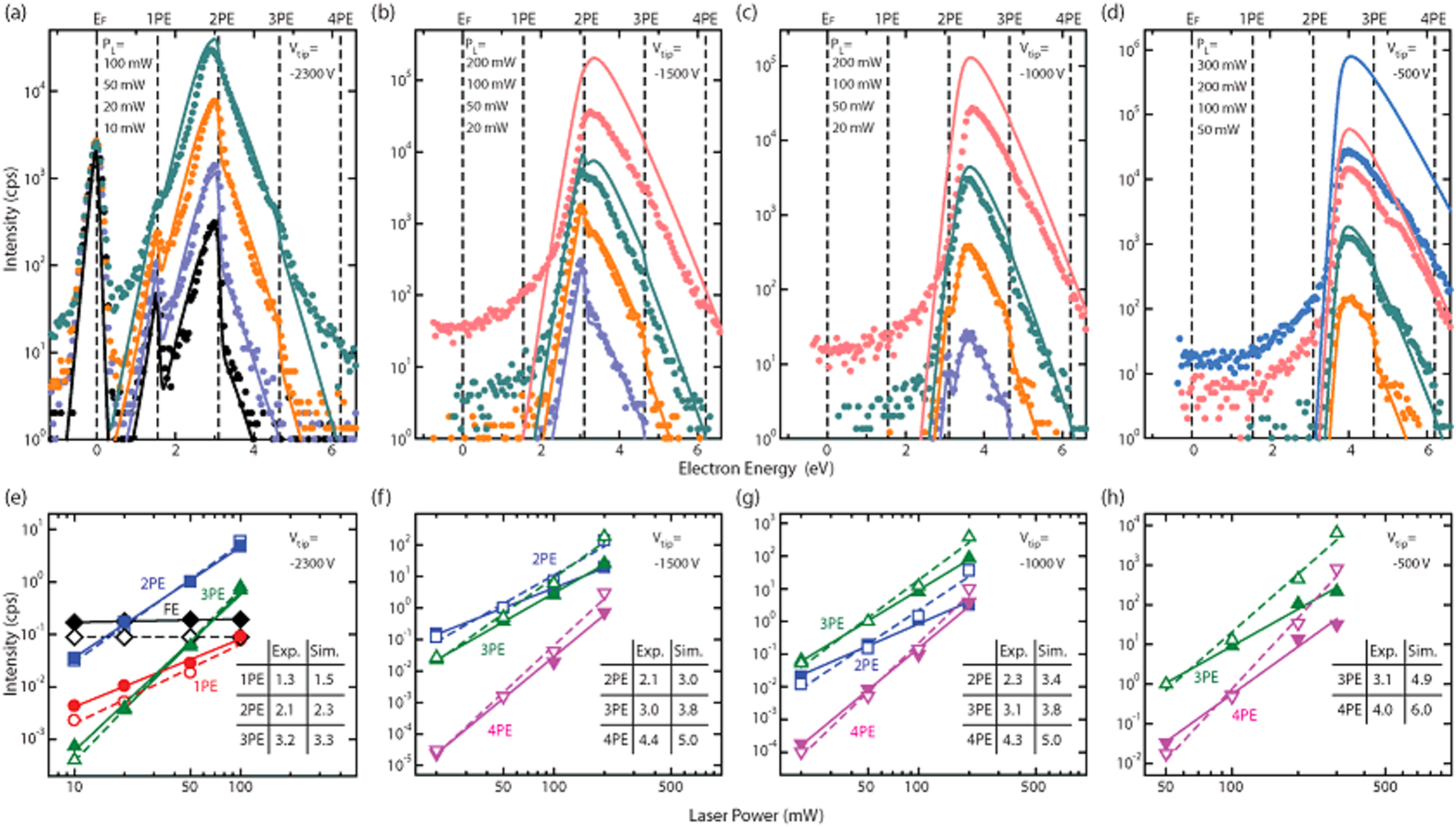}
  \caption{\label{fig:epsart}
Semi logarithmic plots of raw experimental EDCs in Fig. 7: (a) $V_{tip}$=-2300 V, (b) $V_{tip}$=-1500 V, (c) $V_{tip}$=-1000 V, and (d) $V_{tip}$=-500 V. The solid lines indicate the simulations renormalized at maximum intensity in the EDCs at 50 mW of each voltage setting. (e) - (h) show the measured laser power dependence of intensities for each excitation order: (e) $V_{tip}$=-2300 V, (f) $V_{tip}$=-1500 V, (g) $V_{tip}$=-1000 V, and (h) $V_{tip}$=-500 V. These intensities were obtained by integrating intensities over the respective energy regions: field emission (FE) is for -1.2 eV $\sim$ 0.3 eV, photo-assisted field emission from 1PE for 0.3 eV $\sim$ 1.85 eV and 2PE for 1.85 eV $\sim$ 3.4 eV and photoemission from 3PE for 3.4 eV $\sim$ 4.95 eV. The solid symbols indicate the experimental data, while open symbols are from the simulations. The solid lines (experiment) and the dashed lines (simulations) are fitting curves with power functions; their exponents are also shown in the insets.}
\end{figure*}

The laser power dependence of intensities in the EDCs has also examined. Figs. 8(a)-8(d) show the raw data of Fig. 7 as solid circles in a semi-logarithmic plot for each voltage setting. The intensities of the photo-assisted field emission grow with increasing laser power for all voltage settings, while the field emission peak remains constant as shown in Fig. 8(a). From these raw spectra, integrated intensities for each excitation order are extracted as a function of laser power and shown as solid symbols in Figs. 8(e)-8(h). These intensities were obtained by integrating intensities over the respective energy regions: field emission (FE) is for -1.2 eV $\sim$ 0.3 eV, photo-assisted field emission from 1PE for 0.3 eV $\sim$ 1.85 eV and 2PE for 1.85 eV $\sim$ 3.4 eV and photoemission from 3PE for 3.4 eV $\sim$ 4.95 eV. The solid lines are fitted curves with power functions. Exponents for each excitation order are also listed in the insets. These exponents are quite close to the corresponding excitation order, which is consistent with previous work \cite{ropers07, yanagisawa11,barwick07 }.


\subsection{Simulations of EDCs}

The experimental EDCs were reproduced and the influence of electron dynamics on the EDCs was also examined by simulating transient EDCs, taking into account all relevant electron scattering processes, containing both, contributions from new single-photon excitations from the current time step, and from the partly relaxed electron distribution from all previous time steps. A laser pulse with a temporal width of 100 fs was moved across the emission site on the tip apex in 0.2 fs steps. The temporal evolution is covered from -200 fs to 1200 fs for the laser power below 100 mW, while -400 fs to 2400 fs for 200 mW and -400 fs to 3800 fs for 300 mW. The time zero is defined when the maximum of the laser pulse meets the emission site. At each time step, the transient electron distribution function and the resulting EDC were calculated \cite{wu08, rethfeld02,rethfeld99,fatti00,pietanza07}. At the end of the each time window, the transient total yield becomes approximately one percent of its maximum.  All these transient EDCs were integrated over the complete time window. EDCs of field emission without the laser excitation was also calculated for the rest of one period of the laser pulse cycle (approximately 10 ns), and was added to the EDCs for the first laser-induced-emission period. The resulting time-integrated EDCs were normalized at the maximum intensity. Thus obtained simulations were compared with the normalized experimental EDCs in Fig. 7; comparison of the absolute intensities were difficult because of the uncertain transmission of electron beams in the spectrometer. There are only three fitting parameters: 1) the work function $\Phi$, 2) the DC field $F_{DC}$, and 3) the excitation constant $C_{exc}$ which is proportional to the laser power; the definition of $C_{exc}$ is given in the Ref. \cite{yanagisawa11}. More explanations on the simulation is also provided in Ref. \cite{yanagisawa11}. 

The detailed fitting procedure is the following. Firstly, three fitting parameters are adjusted at one particular setting of the tip voltage and laser power ($V_{tip}$=-500 V; $P_{L}$ = 50 mW), and then time-integrated EDCs for the other settings were simulated by scaling up or down $F_{DC}$ and $C_{exc}$ according to the corresponding tip voltage and laser power. This procedure was iterated until reasonable fitting was obtained for all settings. Thus all the experimental EDCs in Fig. 7 were reproduced by such simulations. Throughout all the various conditions, the simulations are in very good accordance with the spectral shapes of the experimental EDCs, as shown by solid lines in Fig. 7. Thus we conclude that the electron dynamics play a significant role in photo-assisted field emission.

Also the intensities of the simulated EDCs were also compared with the experimental data, which are shown as solid lines in Figs. 8(a)-(d). Because it is difficult to simulate absolute intensities, we renormalized all the simulated spectra in Fig. 7 at the maximum intensities of the raw experimental EDCs at 50 mW in each voltage setting. The integrated intensities for each excitation order are also extracted from those simulated spectra and shown as open symbols together with their fitting curves as dashed lines in Figs. 8(e)-(h). The simulations show quite good agreement with experimental data up to 100 mW as can be especially seen in Fig. 8(a) and 8(e). The simulated exponents for each excitation order also show good accordance. Note that these exponents are not exactly 1, 2 and 3, but slightly higher. This is because the higher-order excited electrons decay due to the electron scatterings and some of the decayed electrons are emitted from the lower-order excitation region. 

\begin{figure}[b!]
  \includegraphics[scale=0.24]{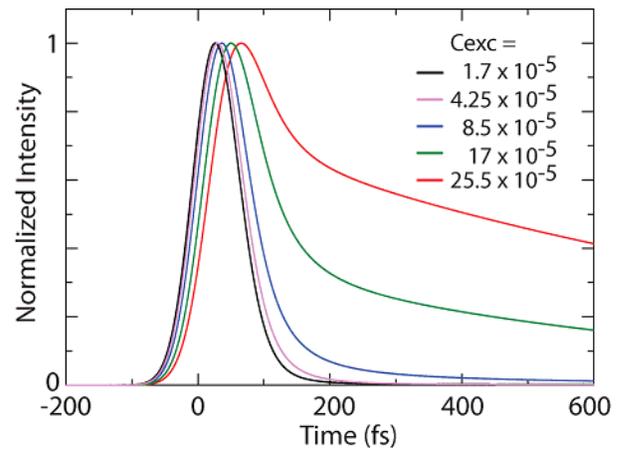}
  \caption{\label{fig:epsart}
Temporal line profiles of the simulated electron pulses for $C_{exc} = 1.7 \times 10^{-5}$, $4.25 \times 10^{-5}$, $8.5 \times 10^{-5}$, $17 \times 10^{-5}$ and $25.5 \times 10^{-5}$. $F_{DC}$ is 0.502 V/nm for all the curves. The spectra are normalized at their maximum.}
\end{figure}

Although simulations show good agreement up to 100 mW, the simulations largely deviate from experimental data at 200 mW and 300 mW. These deviations become bigger with increasing laser power as can be especially seen in Figs. 8(d) and 8(h), and the simulated exponents become higher than the experimental ones due to the inclusion of these data points. Because the experimentally obtained exponents are very close to the corresponding excitation order, we believe our simulation fails for the higher laser power regime. We speculate that the deviation is due to a lack of energy flow to the outside of the system (energy dissipation) in our simulation model. Without the energy dissipation, a decay speed of the excited electrons would be slower than the actual decay  \cite{lisowski04}. As a consequence, the electron emission lasts for a relatively long time, thus leading to an overestimation of the emission intensity. In our simulation, temporal line profiles of the simulated electron pulse as in Fig. 9 show that the electron emission lasts much longer at $C_{exc} = 17 \times 10^{-5}$, $25.5 \times 10^{-5}$ than those at the lower $C_{exc}$ of approximately 200 fs. This indicates that the temperature of the phonon system rises significantly at higher laser power. According to the Ref. \cite{lisowski04}, the inclusion of energy dissipation processes, especially ballistic transport, reduces the heating of the phonon system. Such an energy-dissipation term is necessary for a system where the finite optical penetration depth leads to a spatially inhomogeneous excitation \cite{lisowski04}, which is also the case in the laser-illuminated tip apex \cite{kealhofer12}. Therefore, the inclusion of the energy dissipation in the simulation would solve this problem. Further investigation is required on this point.

\section{Summary}
Illumination of the sharp metallic tip apex with femtosecond laser pulses creates asymmetric local fields and the local field distribution changes with varying laser polarization. Thus, an ultrafast pulsed field-emission source with site selectivity on the scale of a few tens of nanometers was realized. The energy spectra of the laser-induced field emission reveal that the emission mechanism is dominated by photo-assisted field emission in the investigated regime. The emission processes and the spectral shapes can be tuned by the DC field and the laser power. In the higher DC field, photo-assisted field emission is dominant and the emission from lower-excitation order grows with increasing the DC fields. For lower DC fields and higher laser power, photo-assisted field emission from higher-excitation order and photoemission processes become dominant. Together with the simulations, the importance of the electron dynamics during the electron emission is clarified. However, the simulations failed to reproduce the experimental data in terms of the emission intensity in the higher laser power regime due to the neglect of the transport effects. 

A main advantage of the optical control of the field emission sites is that it can manipulate the electron emission within the coherence length and time of the electrons inside the solid, which are roughly 100 nm and 100 fs for tungsten at low temperature \cite{cho04}. Hence, this technique is maybe used to measure the coherence of the electron in the solids \cite{yanagisawa09, yanagisawa10}. Moreover, the clarification of the detailed emission mechanism should be useful to design pulsed electron sources for various applications as mentioned in the introduction.

\section{Acknowledgment}
We thank Prof. J\"{u}rg Osterwalder, Dr. Matthias Hengsberger, Dr. B\"{a}rbel Rethfeld, Prof. Kazuyuki Watanabe and Prof. Hans-Werner Fink for a fruitful discussion. 
This work was supported by the Swiss National Science Foundation through the {\it Ambizione} (grant number PZ00P2\_131701) and the {\it NCCR MUST}, the Japan Society for the Promotion of Science (JSPS), Nishina Memorial Foundation and Kazato Research Foundation.


\begin{thebibliography}{99}
\bibitem{gomer93}R. Gomer, "Field Emission and Field Ionization", (American Institute of Physics, New York, 1993).

\bibitem{fursey03}G. Fursey, "Field Emission in Vacuum Microelectronics", (Kluwer Academic / Plenum Publishers, New York, 2003).  
  

\bibitem{nagaoka98}K. Nagaoka, T. Yamashita, S. Uchiyama, M. Yamada, H. Fujii, and C. Oshima, Nature (London). {\bf 396}, 557 (1998).

\bibitem{oshima02}C. Oshima, K. Mastuda, T. Kona, Y. Mogami, M. Komaki, Y. Murata, T. Yamashita, T. Kuzumaki, and Y. Horiike, Phys. Rev. Lett. {\bf 88}, 038301 (2002).

\bibitem{cho04}B. Cho, T. Ichimura, R. Shimizu, and C. Oshima, Phys. Rev. Lett. {\bf 92}, 246103 (2004).

\bibitem{fink86}H. W. Fink, IBM. J. Res. Develop. {\bf 30}, 460 (1986).

\bibitem{fink90}H. -W. Fink, W. Stocker, and H. Schmid, Phys. Rev. Lett. {\bf 65}, 1204 (1990).

\bibitem{fu01}T-Y. Fu, L-C. Cheng, C. -H. Nien, and T. T. Tsong, Phys. Rev. B {\bf 64}, 113401 (2001).


\bibitem{lee73}M. J. G. Lee, Phys. Rev. Lett. {\bf 30}, 1193 (1973).

\bibitem{boussoukaya87}M. Boussoukaya, H. Bergeret, R. Chehab, B. Leblond, J. Leduff and M. Franco, Nucl. Instr. and Meth. A {\bf 256}, 191 (1987).

\bibitem{lee89}M. J. G. Lee, and E.S. Robins, J. Appl. Phys. {\bf 65}, 1699 (1989).



\bibitem{hommelhoff06a}P. Hommelhoff, Y. Sortais, A. Aghajani-Talesh, and M. A. Kasevich, Phys. Rev. Lett. {\bf 96}, 077401 (2006).

\bibitem{hommelhoff06b}P. Hommelhoff, C. Kealhofer, and M. A. Kasevich, Phys. Rev. Lett. {\bf 97}, 247402 (2006).

\bibitem{yanagisawa09}H. Yanagisawa, C. Hafner, P. Don\'{a}, M. Kl\"{o}ckner, D. Leuenberger, T. Greber, M. Hengsberger, and J. Osterwalder, Phys. Rev. Lett. {\bf 103}, 257603 (2009).

\bibitem{yanagisawa10}H. Yanagisawa, C. Hafner, P. Don\'{a}, M. Kl\"{o}ckner, D. Leuenberger, T. Greber, J. Osterwalder, and M. Hengsberger, Phys. Rev. B {\bf 81}, 115429 (2010).


\bibitem{ropers07}C. Ropers, D. R. Solli, C. P. Schulz, C. Lienau, and T. Elsaesser, Phys. Rev. Lett. {\bf 98}, 043907 (2007).

\bibitem{barwick07}B. Barwick, C. Corder, J. Strohaber, N. Chandler-Smith, C. Uiterwaal, and H. Batelaan, New Journal of Physics {\bf 9}, 142 (2007).

\bibitem{wu08}L. Wu, and L. K. Ang, Phys. Rev. B {\bf 78}, 224112 (2008).

\bibitem{tsujino09}S. Tujino, F. Le Pimpec, J. Raabe, M. Buess, M. Dehler, E. Kirk, J. Gobrecht and A. Wrulich, Appl. Phys. Lett. {\bf 94}, 093508 (2009).

\bibitem{yanagisawa11}H. Yanagisawa, M. Hengsberger, D. Leuenberger, M. Kl\"{o}ckner, C. Hafner, T. Greber, and J. Osterwalder, Phys. Rev. Lett. {\bf 107}, 087601 (2011).


\bibitem{hommelhoff10}M. Schenk, M. Kr\"{u}ger and P. Hommelhoff, Phys. Rev. Lett. {\bf 105}, 257601

\bibitem{ropers11}R. Bormann, M. Gulde, A. Weismann, S. V. Yalunin, and C. Ropers,  Phys. Rev. Lett. {\bf 105}, 147601 （2010）.

\bibitem{hommelhoff11}M. Kr\"{u}ger, M. Schenk and P. Hommelhoff, Nature {\bf 475}, 78 (2011).

\bibitem{hommelhoff12a}M. Kr\"{u}eger, M. Schenk, M. Foerster and P. Hommelhoff, J. Phys. B, At. Mol. Opt. Phys. {\bf 45}, 074006 (2012).

\bibitem{hommelhoff12b}M. Kr\"{u}eger, M. Schenk, P. Hommelhoff, G. Wachter, C. Lemell and J. Burgd\"{o}rfer, New Journal of Physics {\bf 14}, 085019（2012）.

\bibitem{hommelhoff12c}G. Wachter, C. Lemell, J. Burgd\"{o}rfer, M. Schenk, M. Kr\"{u}eger and P. Hommelhoff, Phys. Rev. B {\bf 86}, 035402（2012）.

\bibitem{ropers12}G. Herink, D. R. Solli, M. Gulde, and C. Ropers, Nature {\bf 483}, 190 （2012）.





\bibitem{aeschlimann07}M. Aeschlimann, M. Bauer, D. Bayer, T. Brixner, F. J. Garc\'{i}a de Abajo, W. Pfeiffer, M. Rohmer, C. Spindler, and F. Steeb, Nature {\bf 446}, 301 (2007).

\bibitem{tonomura87}A. Tonomura, Rev. Mod. Phys. {\bf 59}, 639 (1987).


\bibitem{ganter08}R. Ganter, R. Bakker, C. Gough, S. C. Leemann, M. Paraliev, M. Pedrozzi, F. Le Pimpec, V. Schlott, L. Rivkin and A. Wrulich, Phys. Rev. Lett. {\bf 100}, 064801 (2008).




\bibitem{novotny97}L. Novotny, R. X. Bian, and X. S. Xie, Phys. Rev. Lett. {\bf 79}, 645 (1997).

\bibitem{novotny02}L. Novotny, J. Am. Ceram. Soc. {\bf 85}, 1057 (2002).

\bibitem{hecht05}B. Hecht, L. Novotny, "Principles of Nano-Optics", (Cambridge University Press, Cambridge, 2005).


\bibitem{sato80}M. Sato, Phys. Rev. Lett. {\bf 45}, 1856 (1980).

\bibitem{kittel}C. Kittel, "Introduction to Solid State Physics" (John Wiley \& Sons, New York, ed. 8, 2005).

\bibitem{zenneck}F. Yang, J. R. Sambles, and G. W. Bradberry, Phys. Rev. B {\bf 44}, 5855 (1991).

\bibitem{christian90}C. Hafner, "The Generalized Multipole Technique for Computational Electromagnetics" (Artech House Books, Boston, 1990).

\bibitem{christian98}C. Hafner, "MaX-1: A Visual Electromagnetics Platform for PCs" (John Wiley \& Sons, Chichester, 1998).

\bibitem{christian99}C. Hafner, "Post-modern Electromagnetics: Using Intelligent MaXwell Solvers" (John Wiley \& Sons, Chichester, 1999).


\bibitem{tungstenepsilon}CRC Handbook of Chemistry and Physics, edited by D. R. Lide (CRC Press, Boca Raton, FL, 2009), 90th ed.

\bibitem{murphy56}E. L. Murphy, and R. H. Good Jr., Phys. Rev. {\bf 102}, 1464 (1956).

\bibitem{young59}R. D. Young, Phys. Rev. {\bf 113}, 110 (1959).

\bibitem{comment1} Note that Fermi level is not exactly the same energy as that of the field emission peak. However, the peak energy is very close to the Fermi level; their energy difference is several tens of meV in our DC field regime. Hence, we defined the peak energy as Fermi level in this manuscript.



\bibitem{lisowski04}M. Lisowski, P.A. Loukakos, U. Bovensiepen, J. St\"{a}hler, C. Gahl, and M. Wolf, Appl. Phys. A {\bf 78}, 165 (2004).

\bibitem{rethfeld02}B. Rethfeld {\it et al.}, Phys. Rev. B {\bf 65}, 214303 (2002).

\bibitem{comment2} It should be mentioned that the DC fields on the tip apex are not proportional to a voltage difference between $V_{pin}$ and $V_{tip}$ because there is metallic lens holder grounded which is approximately 11 mm  away from the tip and it disturbs the linearity between the applied voltages and the DC fields. This can be avoided by setting $V_{pin}$ grounded and applying voltage only on the tip; the EDCs taken with such a condition are shown in Fig. 7.



\bibitem{rethfeld99}B. Rethfeld, Ph.D. thesis, Technische Universit\"{a}t Braunschweig, 1999.

\bibitem{fatti00}N. Del Fatti, C. Voisin, M. Achermann, S. Tzortzakis, D. Christofilos and F. Vall\'{e}e, Phys. Rev. B {\bf 61}, 16956 (2000).

\bibitem{pietanza07}L. D. Pietanza, G. Colonna, S. Longo and M. Capitelli, Eur. Phys. J. D {\bf 45}, 369 (2007).

\bibitem{kealhofer12}C. Kealhofer, S. M. Foreman, S. Gerlich and M. A. Kasevich, Phys. Rev. B {\bf 86}, 035405 (2012).





\end{thebibliography}
\end{document}